\documentclass{aastex}          
\usepackage{spr-astr-addons}    
\usepackage{epsfig}                              

\begin{document}
%
\title{Quasinormal modes of a black hole with quintessence-like matter and a deficit solid angle: scalar and gravitational perturbations}


\author{Ping Xi}
\affil{Shanghai United Center for Astrophysics(SUCA), Shanghai
Normal University, 100 Guilin Road, Shanghai 200234,China}
\email{xiping@shnu.edu.cn}


\begin{abstract}
In the previous paper (Li, X. Z., Xi, P., Zhai, X. H.: Phys. Lett.
B{\bf666}, 125-130 (2008)), we show the solutions of Einstein
equations with static spherically-symmetric quintessence-like matter
surrounding a global monopole. Furthermore, this monopole become a
black hole with quintessence-like matter and a deficit solid angle
when it is swallowed by an ordinary black hole. We study its
quasinormal modes by WKB method in this paper. The numerical results
show that both the real part of the quasinormal frequencies and the
imaginary part decrease as the state parameter $w$, for scalar and
gravitational perturbations. And we also show variations of
quasinormal frequencies of scalar and gravitational fields via
different $\epsilon$ (deficit solid angel parameter) and different
$\rho_0$ (density of static spherically-symmetric quintessence-like
matter at $r=1$), respectively.
\end{abstract}

\keywords{quasinormal modes; black hole with a deficit solid angle;
quintessence-like matter; WKB method}

%
\section{Introduction}
Current observations
\citep{Hinshaw1-1,Ho1-4,Percival1-3,Kowalski1-2} (cosmic microwave
background, Type Ia Supernovae, baryon acoustic oscillation,
integrated Sachs-Wolfe effect correlations, etc.) show that our
universe is accelerating, which suggest the existence of a spatially
homogeneous and gravitationally repulsive energy component referred
as dark energy. The simplest dark energy candidate is the
cosmological constant $\Lambda$ stemming from energy density of the
vacuum, which corresponds to a fluid with a constant equation of
state $w=-1$. But the theoretical value of vacuum energy density is
much larger than the observed \citep{Weinberg02}. So, various
scalar-field dark energy models
\citep{Peebles3-1,Caldwell3-2,Li3-3,Li3-4,Chiba3-5,Piazza3-6} are
presented, such as quintessence ($-1<w<-\frac{1}{3}$), phantom
($w<-1$), etc.. However, these dynamical dark energy candidates lack
a concrete motivation from fundamental physics. Therefore, the
subject of dark energy is still an attractive topic.

On the other hand, the phase transition in the early universe could
have produced different kinds of topological defects, whose
cosmological implications are very important \citep{Vilenkin04}. The
global monopole, which has divergent mass in flat space-time, is one
of the most interesting defects. When one considers gravity, the
linearly divergent mass of the global monopole has an effect
analogous to that of a deficit solid angle plus a tiny mass at the
origin. It has been shown that this effective mass is actually
negative \cite{Shi4-1,Harari4-2}. However, this monopole become a
black hole with a deficit solid angle (global monopole black hole)
when it is swallowed by an ordinary black hole \cite{Brriola05}.

Recently, black holes enveloped by the quintessence field have been
widely investigated. Kiselev \cite{Kiselev} provided a new black
hole solution using Einstein equations with static
spherically-symmetric quintessence-like matter. Then, quasinormal
modes of this kind of black hole have been studied
\cite{chen,zhang}, which are believed by some physicists to be a
unique fringerprint in directly identifying the existence of a black
hole. In this paper, we show the solution of black hole with
quintessence-like matter and a deficit solid angle, and study its
QNMs by WKB method \cite{Ferrari09}. The numerical results show both
the real part and the imaginary part of the quasinormal frequencies
for scalar and gravitational perturbations decrease as the state
parameter $w$; oscillating frequencies and damping frequencies of
scalar and gravitational fields increase with the deficit solid
angel parameter $\epsilon$ decreasing and density of static
spherically-symmetric quintessence-like matter $\rho_0$ decreasing,
respectively.

\section{Metric, scalar and gravitational perturbations}

We shall work within a particular model in unit $c = 1$, where a
global $O(3)$ symmetry is broken down to $U(1)$. The Lagrangian
density is
\begin{equation}
\mathcal{L}=
\frac{1}{2}g^{\mu\nu}\partial_{\mu}\phi^{a}\partial_{\nu}\phi^{a}-\frac{\lambda^{2}}{4}(\phi^{a}\phi^{a}-{\sigma_{0}}^{2})^{2}.
\end{equation}
where $\phi^{a}$ is triplet of scalar fields, isovector index $a =
1, 2, 3$. The hedgehog configuration describing a global monopole is
\begin{eqnarray}
\phi^{a} = \sigma_{0}f(\tilde{r})\frac{x^{a}}{\tilde{r}},\indent
with \indent x^{a}x^{a}=\tilde{r}^{2}.
\end{eqnarray}
so that we shall actually have a monopole solution if $f \rightarrow
1$ at spatial infinity and $f \rightarrow 0$ near the origin.

The general static metric with spherical symmetry can be written as
\begin{equation}
ds^2=B(\tilde{r})dt^2-A(\tilde{r})dr^2-\tilde{r}^2(d\theta^2+\sin^2\theta
d\phi^2),
\end{equation}
The solutions for a global monopole surrounded by the static
spherically-symmetric quintessence-like matter are as follows
\cite{Li8-1}:
\begin{equation}
\begin{split}
ds^2=&(1-\frac{2G\sigma_0m}{r}-\epsilon^2+\frac{\rho_{0}}{{3w}}r^{-3w-1})dt^2\\
&-\frac{1}{1-\frac{2G\sigma_0m}{r}-\epsilon^2+\frac{\rho_{0}}{{3w}}r^{-3w-1}}dr^2\\
&-r^2(d\theta^2+\sin^2\theta d\phi^2),
\end{split}
\end{equation}
where $\epsilon\equiv\sqrt{8\pi G\sigma_0^2}$ is a dimensionless
parameter of a deficit solid angle, a dimensionless $r\equiv
\sigma_{0}\tilde{r}$, and $m\approx-\frac{16\pi\sigma_0}{3\lambda}$.

When such a global monopole is swallowed by an ordinary black hole
with mass \~{M}, a black hole with quintessence-like matter and a
deficit solid angle can be formed \cite{Brriola05} :
\begin{equation}
\begin{split}
ds^2=&(1-\frac{2M}{r}-\epsilon^2+\frac{\rho_{0}}{{3w}}r^{-3w-1})dt^2\\
&-\frac{1}{1-\frac{2M}{r}-\epsilon^2+\frac{\rho_{0}}{{3w}}r^{-3w-1}}dr^2\\
&-r^2(d\Omega^2),
\end{split}
\end{equation}
where $M=G\sigma_0$(\textit{\~{M}}$-m$) is the dimensionless
parameter of global monopole black hole mass surrounded by
quintessence-like matter.

Now, we consider concretely the behaviors of scalar and
gravitational perturbations in a black hole with quintessence-like
matter and a deficit solid angle, respectively. The propagation of a
massless scalar field is described by the Klein-Gordon equation
\begin{equation}
\Box\Phi=0,
\end{equation}
Then we separate variables by setting
\begin{equation}
\Phi(t,r,\theta, \phi)=\frac{1}{r}\psi(r)Y_{lm}(\theta,
\phi)e^{-i\omega t},
\end{equation}
where $Y_{lm}(\theta, \phi)$ are the usual spherical harmonics.
Submitting Eq.(7) to (6), we obtain
\begin{equation}
\frac{d^2\psi(r)}{dr_\ast^2}+(\omega^2-V_s)\psi(r)=0,
\end{equation}
where $r_\ast$ is the tortoise coordinate
\begin{equation}
r_\ast=\int{\frac{1}{1-\frac{2M}{r}-\epsilon^2+\frac{\rho_{0}}{{3w}}r^{-3w-1}}}dr,
\end{equation}
and $V_s$ is the effective potential
\begin{equation}
\begin{split}
V_s=&({1-\frac{2M}{r}-\epsilon^2+\frac{\rho_{0}}{3w}r^{-3w-1}})\\
&\times[\frac{l(l+1)}{r^2}+\frac{2M}{r^3}+\frac{\rho_{0}(3w+1)}{3w}r^{-3w-3}],
\end{split}
\end{equation}

For gravitational perturbations, the metric function is expressed as
\begin{equation}
g_{\mu\nu}=\bar{g}_{\mu\nu}+h_{\mu\nu},
\end{equation}
where $\bar{g}_{\mu\nu}$ is the background metric, and $h_{\mu\nu}$
is a small perturbation. Here, We adopt the canonical form for
$h_{\mu\nu}$ in classical Regge-Wheeler gauge \cite {Regge}
\begin{displaymath}
h_{\mu\nu}=
\left(\begin{array}{cccc}
0&0&0&h_{0}(r)\\
0&0&0&h_{1}(r)\\
0&0&0&0\\
h_0(r)&h_1(r)&0&h_0(r)
\end{array}\right)e^{-i\omega t}(\sin\theta\frac{\partial}{\partial\theta})P_{l}(\cos\theta),
\end{displaymath}
Introducing
$Q(r)=\frac{1-\frac{2M}{r}-\epsilon^2+\frac{\rho_0}{3w}r^{-3w-1}}{r}h_1(r)$,
we obtain
\begin{equation}
\frac{d^2Q(r)}{dr_\ast^2}+(\omega^2-V_g)Q(r)=0,
\end{equation}
where $V_g$ is the effective potential
\begin{equation}
\begin{split}
V_g=&({1-\frac{2M}{r}-\epsilon^2+\frac{\rho_{0}}{3w}r^{-3w-1}})\\
&\times[\frac{l(l+1)}{r^2}-\frac{6M}{r^3}+\frac{\rho_{0}(3w+1)}{3w}r^{-3w-3}],
\end{split}
\end{equation}

\section{WKB method and numerical results}
Here, we make use of the third-order WKB method to numerical
calculation. The QNMs is as follows
\begin{equation}
\omega^2=[V_{0}+(-2V_{0}^{''})^{\frac{1}{2}}\Delta]-i(n+\frac{1}{2})(-2V_{0}^{''})^\frac{1}{2}(1+\Omega),
\end{equation}
where
\begin{gather}
\Delta=\frac{1}{(-2V^{''}_0)^{1/2}}\nonumber\\\left\{\frac{1}{8}\left(\frac{V^{(4)}_0}{V^{''}_0}\right)
\left(\frac{1}{4}+\alpha^2\right)-\frac{1}{288}\left(\frac{V^{'''}_0}{V^{''}_0}\right)^2
(7+60\alpha^2)\right\},\nonumber\\
\Omega=\frac{1}{(-2V^{''}_0)^{1/2}}\bigg\{\frac{5}{6912}
\left(\frac{V^{'''}_0}{V^{''}_0}\right)^4
(77+188\alpha^2)\nonumber\\-
\frac{1}{384}\left(\frac{V^{'''^2}_0V^{(4)}_0}{V^{''^3}_0}\right)
(51+100\alpha^2)
+\frac{1}{2304}\left(\frac{V^{(4)}_0}{V^{''}_0}\right)^2\nonumber\\(67+68\alpha^2)
+\frac{1}{288}
\left(\frac{V^{'''}_0V^{(5)}_0}{V^{''^2}_0}\right)(19+28\alpha^2)-\frac{1}{288}\nonumber\\
\left(\frac{V^{(6)}_0}{V^{''}_0}\right)(5+4\alpha^2)\bigg\},\nonumber
\end{gather}
and
\begin{eqnarray}
\alpha=n+\frac{1}{2},\;\;\;\;\;
V^{(n)}_0=\frac{d^nV}{dr^n_*}\bigg|_{\;r_*=r_*(r_{p})}\nonumber
\end{eqnarray}

\begin{table*}
\caption{QNMs of global monopole black hole surrounded by static
spherically-symmetric quintessence-like matter with different state
parameter $w$ for scalar perturbations.}
\begin{tabular}{ccccc}
 \hline
 $w$ &$\omega\;(n=0)$& $\omega \;(n=1)$ &$\omega \;(n=2)$ &
 $\omega \;(n=3)$ \\ \hline
 -5/12&0.28580-0.09569i&0.25764-0.30009i&0.21992-0.51424i&0.17134-0.73078i
 \\
-7/12&0.28435-0.09527i&0.25646-0.29868i&0.21907-0.51174i&0.17091-0.72716i
 \\
-8/12&0.28322-0.09503i&0.25563-0.29779i&0.21857-0.51005i&0.17077-0.72464i
 \\
-9/12&0.28177-0.09481i&0.25467-0.29680i&0.21807-0.50801i&0.17075-0.72153i
\\
-11/12&0.27773-0.09477i&0.25287-0.29427i&0.21714-0.50139i&0.16964-0.71093i
\\
\hline
\end{tabular}
\end{table*}

\begin{table*}
\small \caption{QNMs of black hole with a deficit solid angle for
scalar perturbations.}
\begin{tabular}{cccc}
 \hline
 $\omega\;(n=0)$& $\omega \;(n=1)$ &$\omega \;(n=2)$ &
 $\omega \;(n=3)$  \\
 \hline
0.29066-0.09780i& 0.26182-0.30680i&0.22324-0.52574i&
0.17350-0.74710i
 \\
\hline
\end{tabular}
\end{table*}

\begin{table*}
\small \caption{QNMs for scalar perturbations of global monopole
black hole surrounded by static spherically-symmetric
quintessence-like matter with different $\rho_0$ for scalar
perturbations.}
\begin{tabular}{ccccc}
 \hline
 $\rho_0$ &$\omega\;(n=0)$& $\omega \;(n=1)$ &$\omega \;(n=2)$ &
 $\omega \;(n=3)$ \\ \hline
 0.1&0.21180-0.06850i&0.19486-0.21269i&0.17140-0.36264i&0.14100-0.51411i
 \\
0.01&0.28322-0.09503i&0.25563-0.29779i&0.21857-0.51005i&0.17077-0.72464i
 \\
0.001&0.28992-0.09753i&0.26121-0.30591i&0.22277-0.52417i&0.17323-0.74486i
 \\
0.0001&0.29059-0.09778i&0.26176-0.30671i&0.22319-0.52558i&0.17348-0.74687i
\\
\hline
\end{tabular}
\end{table*}

\begin{table*}
\small \caption{QNMs for scalar perturbations of global monopole
black hole surrounded by static spherically-symmetric
quintessence-like matter with different $\epsilon$ for scalar
perturbations.}
\begin{tabular}{ccccc}
 \hline
 $\epsilon^2$ &$\omega\;(n=0)$& $\omega \;(n=1)$ &$\omega \;(n=2)$ &
 $\omega \;(n=3)$ \\ \hline
 0.1&0.23936-0.07645i&0.21773-0.23880i&0.18827-0.40891i&0.15055-0.58106i
 \\
0.01&0.27913-0.09326i&0.25210-0.29216i&0.21577-0.50040i&0.16894-0.71094i
 \\
0.001&0.28322-0.09503i&0.25563-0.29779i&0.21857-0.51005i&0.17077-0.72464i
 \\
0.0001&0.28363-0.09521i&0.25598-0.29836i&0.21885-0.51102i&0.17096-0.72602i
\\
0.00001&0.28368-0.09522i&0.25602-0.29842i&0.21888-0.51112i&0.17098-0.72616i
\\
\hline
\end{tabular}
\end{table*}

\begin{table*}
\small \caption{QNMs of ordinary black hole with quintessence-like
matter for scalar perturbations.}
\begin{tabular}{cccc}
 \hline
 $\omega\;(n=0)$& $\omega \;(n=1)$ &$\omega \;(n=2)$ &
 $\omega \;(n=3)$  \\
 \hline
0.28368-0.09523i&0.25602-0.29842i&0.21888-0.51113i&
0.17098-0.72617i \\
\hline
\end{tabular}
\end{table*}

\begin{table*}
\small \caption{QNMs of global monopole black hole surrounded by
static spherically-symmetric quintessence-like matter with different
state parameter $w$ for gravitational perturbations.}
\begin{tabular}{ccccc}
 \hline
 $w$ &$\omega\;(n=0)$& $\omega \;(n=1)$ &$\omega \;(n=2)$ &
 $\omega \;(n=3)$ \\ \hline
 -5/12&0.32340-0.07088i&0.30367-0.21761i&0.27188-0.37209i&0.23120-0.53110i
 \\
-7/12&0.30746-0.06531i&0.28948-0.20031i&0.26028-0.34243i&0.22295-0.48887i
 \\
-8/12&0.29379-0.06155i&0.27752-0.18841i&0.25074-0.32151i&0.21620-0.45855i
 \\
-9/12&0.27429-0.05714i&0.26090-0.17408i&0.23807-0.29550i&0.20774-0.42002i
\\
-11/12&0.20203-0.04446i&0.19821-0.13362i&0.19091-0.22321i&0.18025-0.31318i
\\
\hline
\end{tabular}
\end{table*}

\begin{table*}
\small \caption{QNMs of global monopole black hole surrounded by
static spherically-symmetric quintessence-like matter with different
$\rho_0$ for gravitational perturbations.}
\begin{tabular}{ccccc}
 \hline
 $\rho_0$ &$\omega\;(n=0)$& $\omega \;(n=1)$ &$\omega \;(n=2)$ &
 $\omega \;(n=3)$ \\ \hline
 0.1&0.29379-0.06155i&0.27752-0.18841i&0.25074-0.32151i&0.21620-0.45855i
 \\
0.01&0.36584-0.08635i&0.33980-0.26596i&0.29838-0.45565i&0.24512-0.65089i
 \\
0.001&0.37203-0.08878i&0.34506-0.27354i&0.30225-0.46868i&0.24713-0.66949i
 \\
0.0001&0.37264-0.08902i&0.34558-0.27429i&0.30263-0.46998i&0.24733-0.67135i
\\
\hline
\end{tabular}
\end{table*}

\begin{table*}
\small \caption{QNMs of global monopole black hole surrounded by
static spherically-symmetric quintessence-like matter with different
$\epsilon^2$ for gravitational perturbations.}
\begin{tabular}{ccccc}
 \hline
 $\epsilon^2$ &$\omega\;(n=0)$& $\omega \;(n=1)$ &$\omega \;(n=2)$ &
 $\omega \;(n=3)$ \\ \hline
 0.1&0.23642-0.04486i&0.22619-0.13670i&0.20891-0.23234i&0.18650-0.33066i
 \\
0.01&0.28870-0.05998i&0.27303-0.18352i&0.24719-0.31306i&0.21387-0.44643i
 \\
0.001&0.29379-0.06155i&0.27752-0.18841i&0.25074-0.32151i&0.21620-0.45855i
 \\
0.0001&0.29379-0.06155i&0.27752-0.18841i&0.25074-0.32151i&0.21620-0.45855i
\\
\hline
\end{tabular}
\end{table*}

\begin{figure}
\center{ \epsfig{file=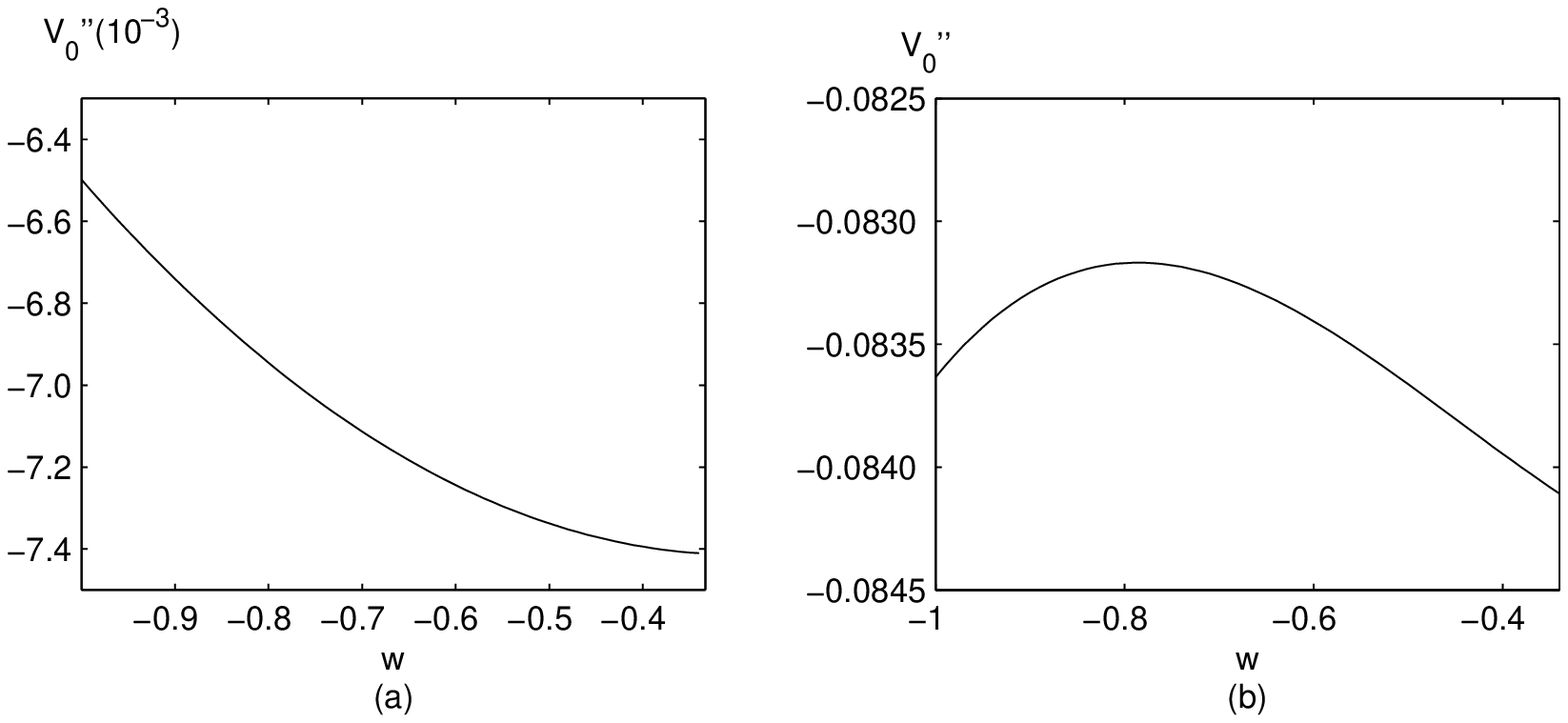,height=2.4in,width=3.2in} \caption{The
second-order derivative of effective potential at its maximum value
$V_0^{\prime\prime}$ are plotted for different values of $w$. The
variation of $V_0^{\prime\prime}$ with $w$ in terms of our numerical
results is shown in (a); the variation of $V_0^{\prime\prime}$ with
$w$ referring to that in Ref.\cite{chen} is shown in (b).}}
\end{figure}

Our numerical results of QNMs for scalar and gravitational
perturbations are listed in Tables 1-5 and Tables 6-8, respectively.
As a reminder, the oscillating quasi-period and the damping time
scale are shown in these tables. In Table 1, we fixed $M=1$,
$\epsilon^2=0.001$, $\rho_0=0.01$ and $l=1$, and the complex
quasinormal frequencies vary with the state parameter $w$.
Obviously, the real parts of the quasinormal frequencies decrease
with $w$, which means the larger the value of $w$ is, more quickly
the global monopole black hole surrounded by quintessence-like
matter oscillates. And the magnitude of imaginary part decreases as
the absolute value of $w$ increases, corresponding to the case more
slowly oscillation of quintessence-like matter surrounding global
monopole black hole decays, smaller the state parameter $w$ is. This
behavior is different from that in Ref.\cite{chen}, in which it was
found that the perturbation damps quicker with the increase of the
absolute value of the equation of state of quintessence. The
difference may be caused by variation of the second-order derivative
of the effective potential at its maximum value with $w$ as shown in
Fig. 1. In Table 2, we chose $M=1$, $\epsilon^2=0.001$ and $l=1$,
and listed quasinormal frequencies of global monopole black hole.
Comparing Table 2 with Table 1, we find that quasi-period of the
oscillation and the imaginary part of quasinormal frequencies in
Table 2 are smaller than that in Table 1. In other words,
oscillation of global monopole black hole decays more slowly in
quintessence-like matter case. In Table 3, we consider the
quasinormal frequencies vary with different $\rho_0$ (density of
static spherically-symmetric quintessence-like matter at $r=1$)
fixing $M=1$, $\epsilon^2=0.001$, $w=-\frac{2}{3}$ and $l=1$. The
oscillating frequency and the damping frequency both increase when
$\rho_0$ decreases. In Table 4, we show how the QNMs behave for
various deficit solid angels in $w=-\frac{2}{3}$, $M=1$,
$\rho_0=0.01$ and $l=1$ case. It is clear that the real part of the
quasinormal frequencies increases when $\epsilon$ decreases.
However, the imaginary part decreases as $\epsilon$. It is
interesting to note that when $\epsilon$ small enough, the real part
of quasinormal frequencies in Table 4 is very close to that of
ordinary black hole with quintessence-like matter in Table 5
(Parameters of this black hole $w, M, \rho_0$, and $l$ are same as
those in Table 4). However, the damping time scales of black hole
with a deficit solid angle (in Table 4) are larger than that of
black hole without deficit solid angle (in Table 5).

For gravitational perturbations, we study variations of quasinormal
frequencies via different $w$, $\rho_0$ and $\epsilon$,
respectively, as showed in Tables 6-8. In Table 6, we choose $M=1$,
$\rho_0=0.1$, $\epsilon^2=0.001$ and $l=2$, and the complex
quasinormal frequencies vary with $w$. The real parts of quasinormal
frequencies decrease with the decrease of $w$, but the imaginary
parts increase. This conclusion is similar to that in
Ref.\cite{zhang}, which means the gravitational perturbations damps
more slowly in quintessence-like matter case. In Table 7, we
considered the relation between quasinormal modes and the parameter
$\rho_0$ with $M=1$, $w=-\frac{2}{3}$, $\epsilon^2=0.001$ and $l=2$.
Evidently, the oscillating quasi-period and damping time scale both
decrease as $\rho_0$. In Table 8, we show the real parts and the
magnitude of the imaginary parts of quasinormal frequencies both
increase as $\epsilon^2$ decrease. To sum up, the variations of
quasinormal frequency with $w$, $\rho_0$ and $\epsilon$ for
gravitational field are similar to those for scalar field.

%
\acknowledgments {This work is supported by National Science
Foundation of China grant No. 10847153.}

\end{document}